\documentclass[11pt]{article}
\usepackage[colorlinks=true, citecolor=blue, urlcolor=blue, linkcolor=blue, breaklinks=true, pdfpagelabels=false]{hyperref}
\usepackage{cite}

\usepackage{hyperref}

\usepackage{amsmath,amsfonts,amssymb}
\usepackage{graphicx}
\usepackage{epsfig,epsf}
\usepackage{bm}
\usepackage{xcolor}

\def\a {\alpha}
\def\b {\beta}

\def\l {\lambda}

\def\bar {\overline}

\def\be {\begin{equation}}
\def\ee {\end{equation}}
\def\bea {\begin{eqnarray}}
\def\eea {\end{eqnarray}}
\def\n {\nonumber}

\def\bra {\langle}
\def\ket {\rangle}

\def\barr{\begin{array}}
\def\earr{\end{array}}

\def\opcit(#1){ {\em op. cit.}, #1}

\def\issue(#1,#2,#3){#1, #2 (#3)} 

\def\equationautorefname~#1\null{Eq.\,(#1)\null}
\def\pageautorefname\nobreakspace{p.}

\makeatletter\renewcommand{\p@subsection}{\thesection.}\makeatother%
 
\begin{document}

\renewcommand*{\thefootnote}{\fnsymbol{footnote}}


\begin{center}
{\Large\bf{Fermionic decay of charged Higgs boson in low mass 
region in Georgi Machacek Model}}


\vspace{5mm}

{\bf Swagata Ghosh}$^{a,b,c}$\footnote{swgtghsh54@gmail.com}

\vspace{3mm}
{\em{${}^a$\ \ \ Department of Physics and Astrophysics, University of 
Delhi, Delhi, India.
}}

{\em{${}^b$\ \ \ Department of Physics, SGTB Khalsa College, University of Delhi, Delhi, India.}}

{\em{${}^c$\ \ \ Department of Physics, Indian Institute of Technology Kharagpur, Kharagpur 721302, India.}}

\end{center}

\begin{abstract}

At the Large Hadron Collider (LHC), ATLAS and CMS collaborations 
observed various decay modes of the light charged Higgs bosons 
produced by top (anti)quark decays. In this paper, we are 
interested in the subsequent decay of the light charged Higgs 
boson into a charm and a strange quark-antiquark pair and 
into a tau and a tau-neutrino pair, separately, in the context 
of the Georgi-Machacek model, which offers a large triplet 
vacuum expectation value (vev) preserving custodial symmetry. 
We show that these experimental observations constrain the 
triplet vev from above. We explore the model parameter space 
consistent with the theoretical constraints, the latest Higgs 
data and the experimental data for light charged Higgs decaying 
to $cs$ and $\tau\nu_{\tau}$.

\end{abstract}



\setcounter{footnote}{0}
\renewcommand*{\thefootnote}{\arabic{footnote}}

\section{Introduction}
\label{intro}

The discovery of the Higgs boson with a mass around 125 GeV 
at the Large Hadron Collider (LHC) experiment 
\cite{ATLAS:2012yve,CMS:2012qbp} in the year of 
2012 let the Standard Model (SM) have great success. The 
possibility of having new exotic particles still cannot be 
thrown away. It is always possible that the discovered 
125 GeV scalar resonance is a part of a non-minimal Higgs 
sector accommodating one or more non-standard scalar 
multiplets. Besides, the new beyond Standard Model (BSM) 
scalar can be neutral or charged. 

With an appealing phenomenology, the physical charged scalars 
got searched at LHC prolongedly. 
These charged scalars can be lighter as well as heavier than 
the top quark mass $(m_t)$. 
At LHC, light charged scalar decaying to $cb$ 
\cite{Ivina:2022tfm, ATLAS:2021zyv}, $cs$
\cite{CMS:2020osd, ATLAS:2013uxj, ATLAS:2010ofa}, 
and $\tau\nu_{\tau}$ \cite{CMS:2019bfg, Abbaspour:2018ysj, ATLAS:2016avi, ATLAS:2011pka, ATLAS:2012nhc, Ali:2011qf, CMS:2012fgz} is searched. 
Besides, light charged scalars decay in different multi-Higgs 
doublet models \cite{Cheung:2022ndq, Akeroyd:2016ymd, Benbrik:2021wyl, Akeroyd:2016ssd, Akeroyd:2012yg, Akeroyd:2022ouy} and in other colliders 
\cite{Hou:2021qff, Akeroyd:2019mvt, Akeroyd:2018axd} 
are also searched and studied.

Most of the BSM models with 
extended scalar sectors commit singly charged scalar particle, 
through which one can explore the model parameter space. In 
this paper, we consider the Georgi-Machacek (GM) Model 
\cite{Georgi:1985nv}, 
where both of the singly and doubly charged scalars are present, 
though we focus only on the decay of the singly charged scalar 
when its mass is in the range of 80-160 GeV. 

The GM model is atypical of the other triplet extensions of 
the SM as it preserves the custodial symmetry in the tree level, 
i.e., the $\rho$-parameter is equal to unity in this model, 
hence resulting in a sizeable triplet vacuum expectation value 
(VEV), $v_2$. In the GM model, the extended scalar sector 
contains one real triplet and one complex triplet, neutral 
components of both possessing the same VEV. The physical scalar 
sector includes ten scalars ordered as two singlets, one 
triplet, and one quintuplet. Out of these ten scalars, 4 scalars 
are charge-neutral ($h$, $H$, $H_3^0$, $H_5^0$), another 4 are 
singly-charged ($H_3^{\pm}$, $H_5^{\pm}$), and the rest 
2 scalars are doubly-charged ($H_5^{\pm\pm}$). The two 
singly-charged scalars $H_5^{\pm}$ do not couple to the SM 
fermions, whereas, the other two singly-charged scalars 
$H_3^{\pm}$ couple to the SM fermions, and the strength of 
coupling is directly proportional to the triplet VEV $v_2$, 
which is sizeable enough in the GM model. We will focus on the 
decay of the $H_3^{\pm}$ in this paper.

Decay of charged scalar, heavier than the top quark mass, is 
studied in the context of the Georgi-Machacek model \cite{Ghosh:2019qie, Logan:2018wtm}. 
The theoretical limits \cite{Ismail:2020zoz, Hartling:2014zca, Krauss:2017xpj, Moultaka:2020dmb, Aoki:2007ah} 
on the Lagrangian 
parameters, the indirect constraints \cite{Hartling:2014aga}, 
Higgs boson pair productions at the LHC \cite{Chang:2017niy}, 
electroweak phase transition \cite{Chiang:2014hia, Zhou:2018zli}, 
impacts of higher dimensional operators \cite{Banerjee:2019gmr} 
in the Georgi-Machacek model are well scrutinized. 
For phenomenological studies of this model at the LHC 
\cite{Chiang:2014bia}, 
electron-positron collider \cite{Chiang:2015rva}, 
the reader may also consult the literature. 
There are also some variants of the GM model \cite{Logan:2015xpa, Kundu:2021pcg}. One can scan the parameter space of this model 
using the calculator \cite{Hartling:2014xma}. For the global 
fits in the GM model, see \cite{Chiang:2018cgb}. 
It is to be mentioned here that, inspite of the extensive 
phenomenological study of the GM model, the exploration of 
the parameter space in case of the light charged scalar 
together with the LHC data has not been examined as yet. 

This paper is organized as follows. 
In Sec. 2 we describe the model in brief. 
We enlist the theoretical constraints and LHC data, with their 
effects on the parameter space of the GM model in Sec. 3. 
The results are given in Sec. 4. 
Finally we conclude in Sec. 5.
%
\section{The Georgi-Machacek Model}
\label{model}

One $SU(2)_L$ real triplet $\left(\xi^{+},\xi^0,\xi^{-}\right)^T$ 
with hyperchage $Y=0$ and one $SU(2)_L$ complex triplet 
$\left(\chi^{++},\chi^{+},\chi^0\right)^T$ with $Y = 2$ 
are appended to the particle contents of the scalar sector of 
the Standard Model to obtain the Georgi-Machacek model. 
We mostly persue the notations followed in 
\cite{Hartling:2014zca} throughout this paper. 

The SM doublet $\left(\phi^{+},\phi^{0}\right)^T$ with $Y = 1$ 
and the non-standard triplets can be expressed in terms of 
bi-doublet and bi-triplet respectively, as, 

\be
\Phi =
\begin{pmatrix}
 \phi^{0*} & \phi^{+}\cr
 \phi^{-} & \phi^0
\end{pmatrix}\,
,\quad
X =
\begin{pmatrix}
 \chi^{0*} & \xi^{+} & \chi^{++}\cr
 \chi^{-} & \xi^0 & \chi^{+}\cr
 \chi^{--} & \xi^{-} & \chi^0
\end{pmatrix}\,.
\ee

The most general scalar potential is given by,  
\bea
V\left(\Phi,X\right) &=& \frac{{\mu_2}^2}{2}\, {\rm Tr}\left(\Phi^\dag\Phi\right) 
+ \frac{{\mu_3}^2}{2}\, \rm{Tr}\left(X^\dag X\right)
+ {\l_1}\left[{\rm Tr}\left(\Phi^\dag\Phi\right)\right]^2  
+ {\l_2}\, {\rm Tr}\left(\Phi^\dag\Phi\right)\, {\rm Tr}\left(X^\dag X\right) \n\\
&& + {\l_3}\, {\rm Tr}\left(X^\dag X X^\dag X\right)
 + {\l_4}\left[{\rm Tr}\left(X^\dag X\right)\right]^2 
- {\l_5}\, {\rm Tr}\left(\Phi^\dag \tau^a \Phi\tau^b\right)\, {\rm Tr}\left(X^\dag t^a X t^b\right) \n\\
&& - {M_1} \, {\rm Tr}\left(\Phi^\dag \tau^a \Phi\tau^b\right) {\left(U X U^\dag\right)_{ab}} 
- {M_2} \, {\rm Tr}\left(X^\dag t^a X t^b\right) {\left(U X U^\dag\right)_{ab}}\, ,
\label{eq:genPot} 
\eea
with $\tau^a = \sigma^a/2$, where $\sigma^a$ are the three Pauli matrices, and the $t^a$s as,
\be
t^1=\frac{1}{\sqrt2}
\begin{pmatrix}
 0 & 1 & 0\cr
 1 & 0 & 1\cr
 0 & 1 & 0
\end{pmatrix}
\,,
t^2=\frac{1}{\sqrt2}
\begin{pmatrix}
 0 & -i & 0\cr
 i & 0 & -i\cr
 0 & i & 0
\end{pmatrix}
\,,
t^3=
\begin{pmatrix}
 1 & 0 & 0\cr
 0 & 0 & 0\cr
 0 & 0 & -1
\end{pmatrix}\,.
\ee
The matrix $U$ in the trilinear terms of the GM potential is 
given by,
\be
U=
\frac{1}{\sqrt{2}}\begin{pmatrix}
 -1 & 0 & 1\cr
 -i & 0 & -i \cr
 0 & \sqrt{2} & 0
\end{pmatrix}\,.
\ee
 
After the electroweak symmetry breaking (EWSB), the neutral 
components of the bi-doublet and the bi-triplet acquire the 
VEVs, as,
\be
\bra\phi^0\ket = \frac{v_1}{\sqrt{2}}\,,\ \ \bra \chi^0 \ket = \bra \xi^0 \ket = {v_2}\,.
\label{eq:vevs}
\ee

Note that, the equality of the real and complex triplet VEVs 
corresponds to the preservation of the custodial symmetry of 
the potential at the tree level leading to 
$\rho_{\rm tree}\, \equiv M_W^2/M_Z^2 \cos^2{\theta_W} = 1$.

The EW VEV $v$ relates to the doublet and triplet VEVs as,
\be
\sqrt{{v_1}^2+8{v_2}^2}=v\approx246~\text{GeV}\,.
\label{eq:v1v2}
\ee

The doublet-triplet mixing angle may be expressed as,
\be
\tan {\b} = \frac{2\sqrt{2}{v_2}}{v_1}\,,
\ee

In terms of VEVs given in Eq.\ (\ref{eq:vevs}), the potential 
(\ref{eq:genPot}) becomes 
\bea
V\left({v_1},{v_2}\right)&=&\frac{{\mu_2}^2}{2}{v_1}^2+3 \frac{{\mu_3}^2}{2}{v_2}^2+{\l_1}{v_1}^4
+ \frac32 \left(2{\l_2}-{\l_5}\right){v_1}^2{v_2}^2\, \n\\
&& +3\left({\l_3}+3{\l_4}\right){v_2}^4
-\frac34{M_1}{v_1}^2{v_2}-6{M_2}{v_2}^3\,,
\label{eq:vevgenPot}
\eea
  
Hence, using the extremisation conditions one can easily extract 
the bilinear coefficients of the potential as,
\bea
{\mu_2}^2=-4{\l_1}{v_1}^2-3\left(2{\l_2}-{\l_5}\right){v_2}^2
+\frac32{M_1}{v_2} \,,\n\\
{\mu_3}^2=-\left(2{\l_2}-{\l_5}\right){v_1}^2-4\left({\l_3}+3{\l_4}\right){v_2}^2
+\frac{M_1 {v_1}^2}{4 v_2}+6{M_2}{v_2} \,.
\label{eq:mu}
\eea

The scalar sector of the GM model consists of ten physical 
fields of which five are part of a custodial quintuplet 
${H_5}=\left(H_5^{++},H_5^{+},H_5^0,H_5^{-},H_5^{--}\right)^T$, 
three are part of a custodial triplet 
${H_3}=\left(H_3^{+},H_3^0,H_3^{-}\right)^T$, and the rest two 
are custodial singlets $H_1^0$ and $H_1^{0'}$. 
These ten physical fields may be expressed in terms of the 
component fields and the angle $\beta$ as,
\bea
&&H_5^{++}=\chi^{++}\,, \ \ \ 
H_5^{+}=\frac{\left(\chi^{+}-\xi^{+}\right)}{\sqrt{2}}\,, \ \ \ 
H_5^0=\sqrt{\frac23}\xi^0-\sqrt{\frac13}\chi^{0R}\,, \n\\
&&H_3^{+}=-\sin {\b}\, \phi^{+}+\cos {\b}\, \frac{\left(\chi^{+}+\xi^{+}\right)}{\sqrt{2}}\,, \ \ \ 
H_3^0=-\sin {\b}\, \phi^{0I}+\cos {\b}\, \chi^{0I}\,, \n\\
&&H_1^0=\phi^{0R}\,, \ \ \ 
H_1^{0'}=\sqrt{\frac13}\xi^0+\sqrt{\frac23}\chi^{0R}\,.
\label{eq:fields}
\eea

The physical fields of the multiplets are mass degenerate. 
The square of the common mass of the quintuplet $H_5$ and 
the triplet $H_3$ are given by,
\bea
{m_5}^2&=&\frac{M_1}{4{v_2}}{v_1}^2+12{M_2}{v_2}+\frac32{\l_5}{v_1}^2+8{\l_3}{v_2}^2\,, \n\\
{m_3}^2&=&\left(\frac{M_1}{4{v_2}}+\frac{\l_5}{2}\right)v^2\,,
\label{eq:m3m5}
\eea
respectively.

The custodial singlets $H_1^0$ and $H_1^{0'}$ mix to yield 
the mass eigenstates $h$ and $H$ as,
\bea
h &=& \cos {\a}\,\, H_1^0-\sin {\a}\,\, H_1^{0'}\,, \n\\ 
H &=& \sin {\a}\,\, H_1^0+\cos {\a}\,\, H_1^{0'}\,,
\label{eq:mhmH}
\eea
with the mass squared matrix
\be
{\cal M}^2=
\begin{pmatrix}
 {{\cal M}_{11}}^2 & {{\cal M}_{12}}^2\cr
 {{\cal M}_{21}}^2 & {{\cal M}_{22}}^2
\end{pmatrix}\,,
\ee
where
\bea
{\cal M}_{11}^2 &=& 8 {\l_1}{v_1}^2\,, \n\\
{\cal M}_{12}^2 &=& {\cal M}_{21}^2 = \frac{\sqrt{3}}{2}\left[-M_1+4\left(2\l_2-\l_5\right)v_2\right]v_1\,, \n\\
{\cal M}_{22}^2 &=& 
\frac{M_1 v_1^2}{4 v_2} - 6 M_2 v_2 + 8 (\lambda_3 + 3 \lambda_4) v_2^2\,.
\eea
The neutral scalar mixing angle $\alpha$ can be parametrised 
in terms of the components of the mass-squared matrix as,
\be
\tan{2\a}=\frac{2 {{\cal M}_{12}}^2}{{{\cal M}_{22}}^2-{{\cal M}_{11}}^2}\,.
\ee
The mass-squared of the physical neutral scalars are,
\be
{m_{h,H}}^2=\frac12 \left[{{\cal M}_{11}}^2+{{\cal M}_{22}}^2\mp\sqrt{\left({{\cal M}_{11}}^2-
{{\cal M}_{22}}^2\right)^2
+4\left({{\cal M}_{12}}^2\right)^2}\right]\,.
\ee
We assume here that $m_H > m_h$ and consider the CP-even neutral 
scalar with smaller mass as the SM-like Higgs boson, such that 
$m_h\approx125~\text{GeV}$.

The bilinear couplings ($\mu_2^2$ and $\mu_3^2$) present in 
the scalar potential are alredy traded in terms of the VEVs 
$v_1$ and $v_2$ in the Eq.\ (\ref{eq:mu}).
The quartic couplings $\l_i$ ($i=1$ to 5) in the potential 
(\ref{eq:genPot}) can be expressed in terms of the four physical 
masses, $m_h$, $m_H$, $m_3$, $m_5$, and the mixing angle $\a$ 
as,
\bea
\l_1 &=& \frac{1}{8 v_1^2}\left(m_h^2\,\cos^2{\a} + m_H^2\,\sin^2{\a}\right)\,, \n\\
\l_2 &=& 
\frac{1}{8 v_2}\left[\left(m_H^2-m_h^2\right)\,\frac{1}{\sqrt{3}v_1}\sin{2\a}
-M_1 +8 m_3^2 \frac{v_2}{v^2} \right]\,, \n\\
\l_3 &=& \frac{1}{8 v_2^2} \left( m_5^2 -3 m_3^2 \frac{v_1^2}{v^2} 
+\frac{M_1 v_1^2}{2 v_2} - 12 M_2 v_2 \right)\,, \n\\
\l_4 &=& \frac{1}{24 v_2^2} \left( m_h^2\,\sin^2 {\a} + m_H^2\,\cos^2 {\a} - m_5^2
+ 3 m_3^2 \frac{v_1^2}{v^2} - 3 M_1 \frac{v_1^2}{4 v_2} + 18 M_2 v_2 \right)\,, \n\\
\l_5 &=& 2 \left( \frac{m_3^2}{v^2} - \frac{M_1}{4 v_2} \right)\,.
\label{eq:lamda}
\eea
%
\section{Theoretical constraints and LHC data}
\label{constraints}
%
The theoretical constraints, mainly resulting from the 
perturbative unitarity and electroweak vacuum stability, and 
the experimental constraints, typically 
following the LHC data put the limit on the parameters of 
the GM potential. 
Here, we write down the theoretical constraints already 
available in the literature. 
The reader may follow \cite{Hartling:2014zca, Hartling:2014aga} 
for the detailed knowledges of the theoretical constraints. 

Constraints obtained from perturbative unitarity are given by,
\bea
\sqrt{\left( 6\l_1-7\l_3-11\l_4 \right)^2+36\l_2^2}+\mid 6\l_1+7\l_3+11\l_4\mid &<& 4\pi\,, \n\\
\sqrt{\left( 2\l_1+\l_3-2\l_4 \right)^2+\l_5^2}+\mid 2\l_1-\l_3+2\l_4\mid &<& 4\pi\,, \n\\
\mid 2\l_3+\l_4\mid &<& \pi\,, \n\\
\mid \l_2-\l_5\mid &<& 2\pi\,,
\label{eq:unitarity}
\eea
and the constraints obtained from electroweak vacuum stability 
are given by,
\bea
\l_1\, &>&\, 0\,,\n\\
\l_2+\l_3\,&>&\,0\,,\n\\
\l_2+\frac12\l_3\,&>&\,0\,,\n\\
-\mid\l_4\mid +2\sqrt{\l_1\left(\l_2+\l_3\right)}\,&>&\,0\,,\n\\
\l_4-\frac14 \mid\l_5\mid +\sqrt{2\l_1\left(2\l_2+\l_3\right)}\,&>&\,0\,.
\label{eq:stability}
\eea

For the constraints coming from the LHC data, we consider 
the Higgs data at $\sqrt{s}=$13 TeV 
\cite{CMS:2018lkl,ATLAS:2019slw}. Here, we contemplate the 
lightest CP-even scalar $h$ as the SM-like Higgs with mass 
$m_h\approx125$ GeV. Therefore, we require to know the couplings 
of $h$ with the SM fermions and the vector bosons, which we 
enlist below :
\be
g_{h{\bar{f}}f} = -i\frac{m_f}{v}\frac{c_{\a}}{c_{\b}}\,,\quad
i g_{hWW}\eta_{\mu\nu} = i c_W^2 g_{hZZ}\eta_{\mu\nu} = -i \frac{e^2}{6 s_W^2} (8\sqrt{3}s_{\a}v_2\,-\,
3 c_{\a}v_1) \eta_{\mu\nu}\,.
\ee
The ratio of the fermionic and bosonic couplings with $h$ in 
the GM model to that in the SM model are given by,
\be
\kappa_f^h = \frac{v}{v_1}c_{\a}\,,\quad
\kappa_V^h = -\frac{1}{3v}(8\sqrt{3}s_{\a}v_2\,-\,3 c_{\a}v_1)\,.
\ee
For the decay $h\rightarrow\gamma\gamma$, the charged scalars 
coming from the triplet and quintuplet contribute at loop level. 
So, we need the corresponding couplings :
\bea
-i g_{h{H_3^+}{H_3^-}} &=&
-i (64\l_1 c_{\a}\frac{v_2^2 v_1}{v^2}\,-\frac{8}{\sqrt{3}}
\frac{v_1^2 v_2}{v^2}s_{\a}(\l_3+3\l_4)\,- \frac{4}{\sqrt{3}} \frac{v_2 M_1}{v^2} (s_{\a}v_2-\sqrt{3}c_{\a}v_1)\, \n\\
&&-\frac{16}{\sqrt{3}}\frac{v_2^3}{v^2}s_{\a} (6\l_2+\l_5)\, -\,c_{\a}\frac{v_1^3}{v^2}(\l_5-4\l_2)\,+\,
2\sqrt{3}M_2\frac{v_1^2}{v^2}s_{\a}\, \n\\
&& -\,\frac{8}{\sqrt{3}}\l_5\frac{v_1 v_2}{v^2} (s_{\a}v_1\,-\,\sqrt{3}c_{\a}v_2))\,, \n \\
-i g_{h{H_5^+}{H_5^-}} &=& -i g_{h{H_5^{++}}{H_5^{--}}} = 
-i( -8\sqrt{3}(\l_3+\l_4) s_{\a} v_2\,+\,(4\l_2+\l_5)c_{\a}v_1
-\,2\sqrt{3}M_2 s_{\a} )\,.
\label{eq:chargedcoupling}
\eea
One can replace the $\l_i$s present in the Eq.\ (\ref{eq:chargedcoupling}) with the help of the expressions given 
in (\ref{eq:lamda}) to get the couplings in terms of the physical 
masses, mixing angle, VEVs, and the trilinear coefficients. 

The vertices involving the charged scalar $H_3^{\pm}$ and two fermions are given by : 
\bea
 H_3^+\overline{u}d : - i \sqrt{2}\,V_{ud}\, \tan{\beta}\, (\frac{m_u}{v}P_L - \frac{m_d}{v}P_R)&,&
 \quad
 H_3^-\overline{d}u : - i \sqrt{2}\,V_{ud}^*\, \tan{\beta}\, (\frac{m_u}{v}P_R - \frac{m_d}{v}P_L),
 \n\\
 H_3^+\overline{\nu}l : i \sqrt{2}\tan{\beta}\, \frac{m_l}{v}P_R &,&
 \quad
 H_3^-\overline{l}\nu : i \sqrt{2}\tan{\beta}\, \frac{m_l}{v}P_L.
\label{eq:H3pmff}
\eea
Here, $V_{ud}$ is the CKM matrix element and the projection operators $P_{R,L}$ are given by $(1 \pm \gamma_5)/2$.

Considering the Eq.\ (\ref{eq:mu}) and (\ref{eq:lamda}), we 
have nine independent parameters ($m_h$, $m_H$, $m_3$, $m_5$, 
$\sin{\a}$, $v_2$, $v$, $M_1$, $M_2$) and we choose $m_h$ as 
identical to the SM-like Higgs mass.
To scan the parameter space, we choose the mass of the heavier 
CP-even neutral scalar $H$ starting from $150$ GeV and that 
of the members of the multiplets starting from $80$ GeV \cite{ALEPH:2013htx}. 
The upper limit of $m_H$ and $m_5$ are set to $1000$ GeV. 
We vary the triplet VEV $v_2$ in between $0-60$ GeV. 
$v$ is the electroweak VEV as usual. 
Therefore, after setting $m_h$ and $v$ at the fixed values, we are left with seven independent parameters. 
Since our purpose is to study the low mass phenomenology of 
$H_3^+$, we consider $m_3$ in between $80-160$ GeV. 
Finally, we consider $M_1$ and $M_2$ in 
between $[-1000,1000]$ GeV. 

\begin{figure}
 \begin{center}
 \includegraphics[width= 7.5cm]{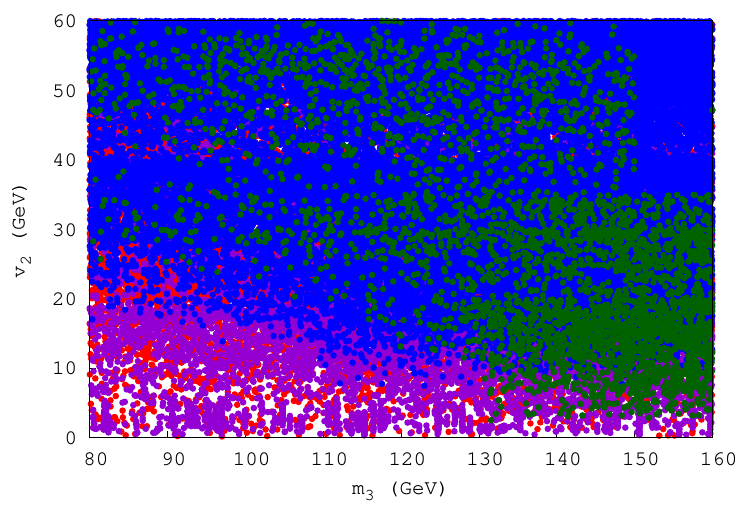} \ \ 
  \includegraphics[width= 7.5cm]{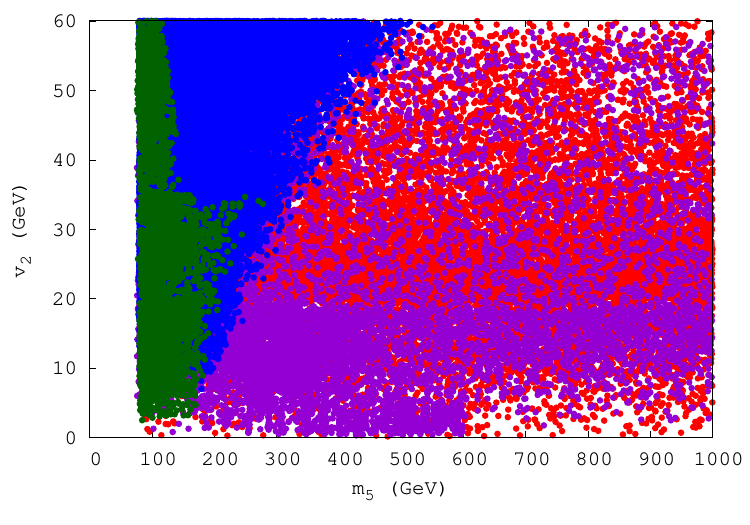} \\
 \includegraphics[width= 7.5cm]{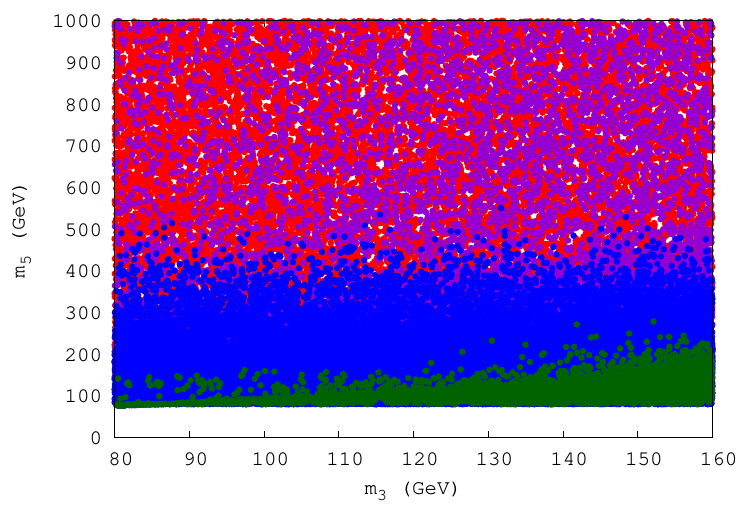} \ \ 
  \includegraphics[width= 7.5cm]{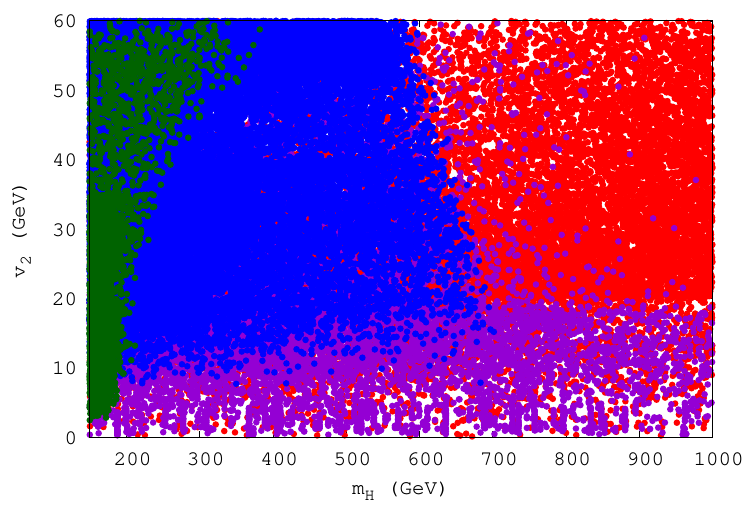} \\
 \includegraphics[width= 7.5cm]{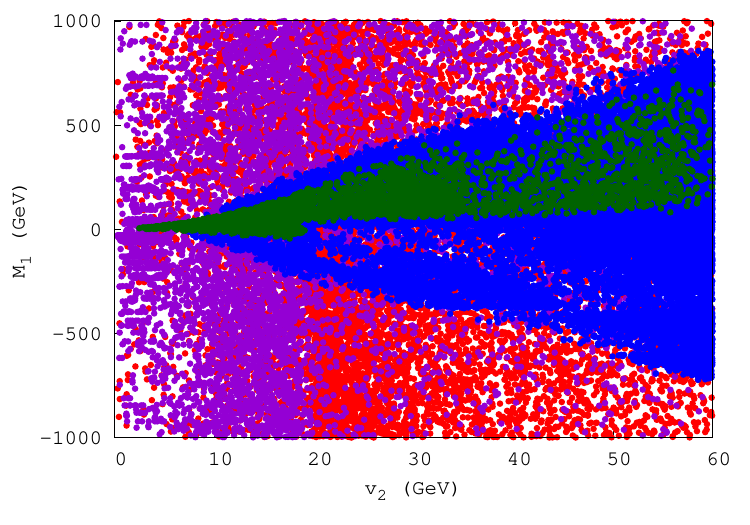} \ \ 
  \includegraphics[width= 7.5cm]{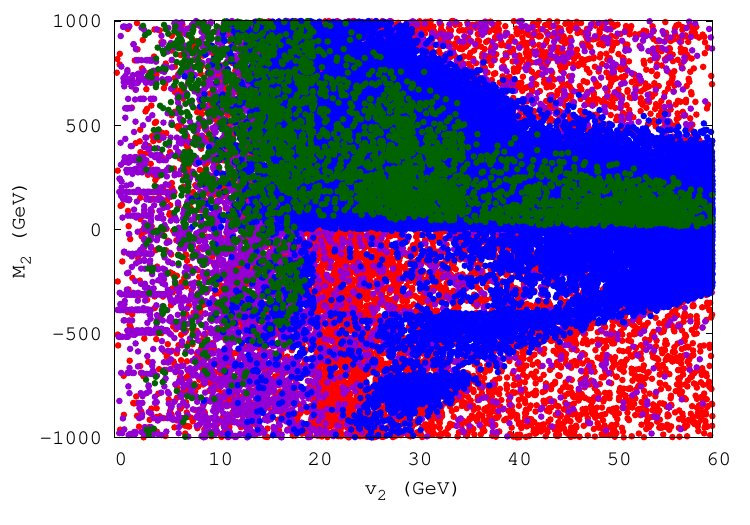} \\
 \includegraphics[width= 7.5cm]{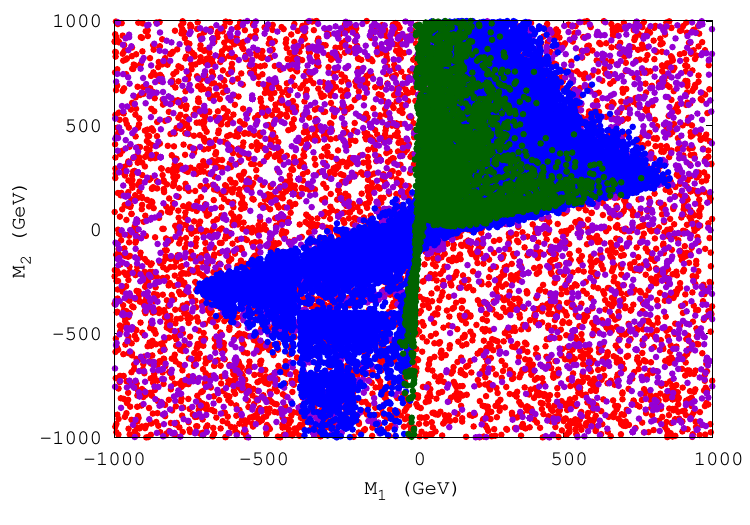} \ \ 
  \includegraphics[width= 7.5cm]{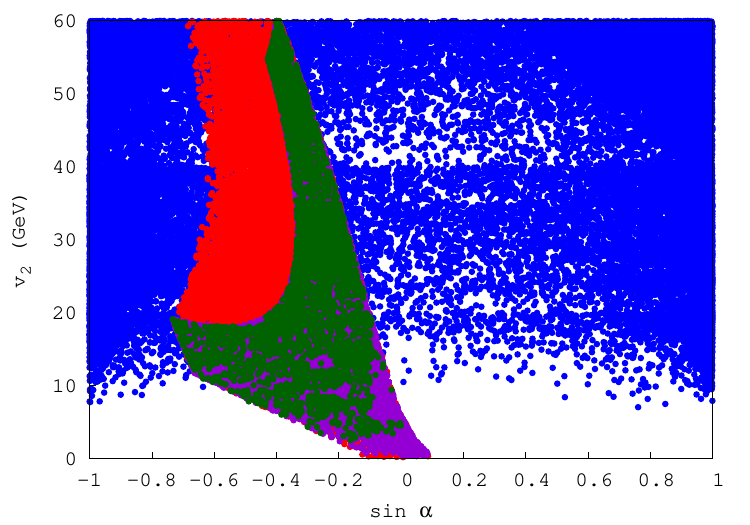} \\
 \end{center}
 \caption{\small Allowed parameter space in the $m_3-v_2$, $m_5-v_2$, $m_3-m_5$, $m_H-v_2$, $v_2-M_1$, $v_2-M_2$, $M_1-M_2$, $\sin{\a}-v_2$ plane by the theoretical constraints, LHC Higgs signal data at $\sqrt{s}=13$ TeV without considering the charged Higgs contribution in the $h \rightarrow \gamma \gamma$ decay, LHC Higgs signal data at $\sqrt{s}=13$ TeV considering the charged Higgs contribution in the $h \rightarrow \gamma \gamma$ decay, all the above stated constraints are shown by blue, red, violet, dark-green points respectively.}
 \label{fig:overlapped}
 \end{figure}

Now, in Fig.\ (\ref{fig:overlapped}) we present the plots 
in the $m_3-v_2$, $m_5-v_2$, $m_3-m_5$, $m_H-v_2$, $v_2-M_1$, 
$v_2-M_2$, $M_1-M_2$, $\sin{\a}-v_2$ plane, where the region 
allowed by the theoretical constraints, LHC Higgs signal data at $\sqrt{s}=13$ TeV without considering the charged Higgs contribution in the $h \rightarrow \gamma \gamma$ decay, LHC Higgs signal data at $\sqrt{s}=13$ TeV considering the charged Higgs contribution in the $h \rightarrow \gamma \gamma$ decay, all the above stated constraints are shown by blue, red, violet, dark-green points respectively. 

\begin{figure}
 \begin{center}
 \includegraphics[width= 9.5cm]{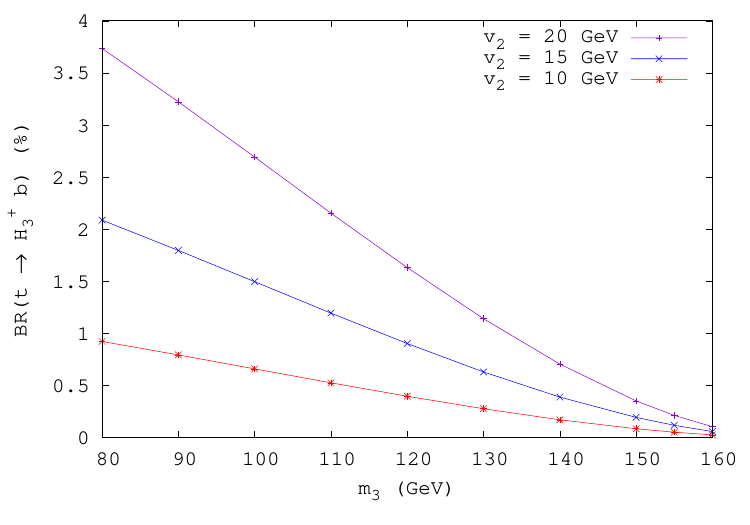}
 \end{center}
 \caption{\small Branching ratio of $t\rightarrow H_3^{+}b$ 
 as a function of mass $m_3$ in GeV, for $v_2=$10,15,20 GeV. 
 The other parameters chosen as mentioned in the Eq.\ (\ref{eq:BP}).}
 \label{fig:BRtTOh3pb}
 \end{figure}

\begin{figure}
 \begin{center}
 \includegraphics[width= 9.5cm]{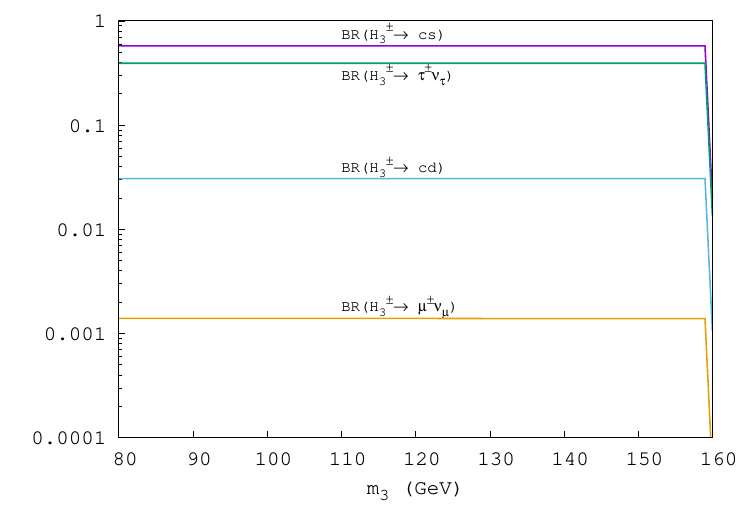}
 \end{center}
 \caption{\small Branching ratios of $H_3^{\pm}$ as a function 
 of mass $m_3$ in GeV at $v_2=20$ GeV. The other independent 
 parameters are chosen as mentioned in the Eq.\ (\ref{eq:BP}).}
 \label{fig:h3pBR}
 \end{figure}

\begin{figure}
 \begin{center}
 \includegraphics[width= 8.2cm]{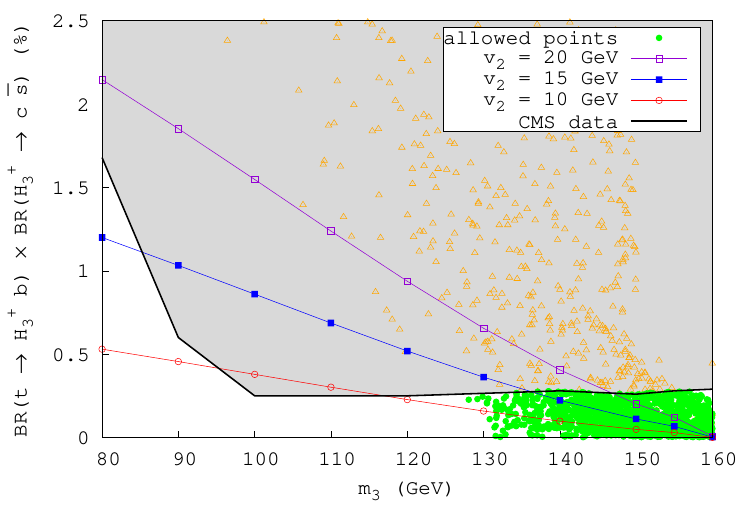} \ \
 \includegraphics[width= 8.2cm]{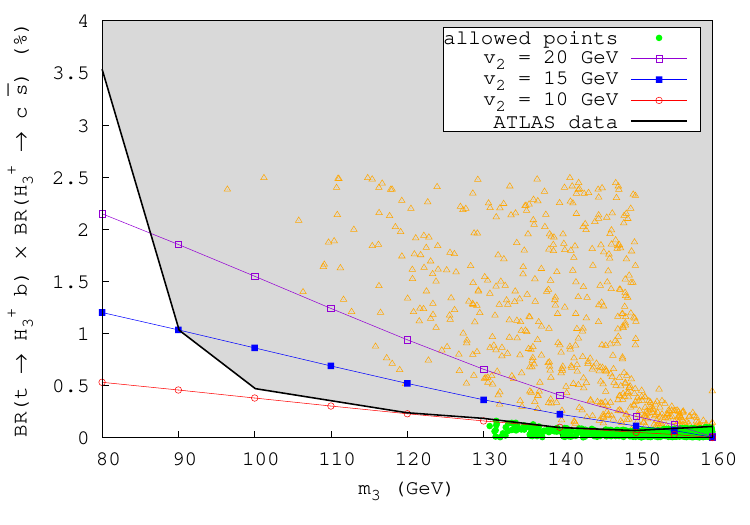}
 \end{center}
 \caption{\small Branching ratio of the top quark decaying into 
 $H_3^+$ and $b$ times the branching ratio of $H_3^+$ 
 decaying into $c$ and $\bar{s}$ as a function of mass $m_3$ in 
 GeV, for $v_2=10,15,20$ GeV with red, blue and violet coloured 
 lines respectively. The black line represents the limit from (Left) CMS experiment \cite{CMS:2020osd} and (Right) ATLAS experiment \cite{ATLAS:2024oqu}. The shaded region is 
 excluded by the experimental data. 
 All the coloured points (green and orange) are allowed by the theoretical constraints. 
 The orange and green points are forbidden and allowed by the corresponding experimental data respectively. }
 \label{fig:cs}
 \end{figure}

\begin{figure}
 \begin{center}
 \includegraphics[width= 8.2cm]{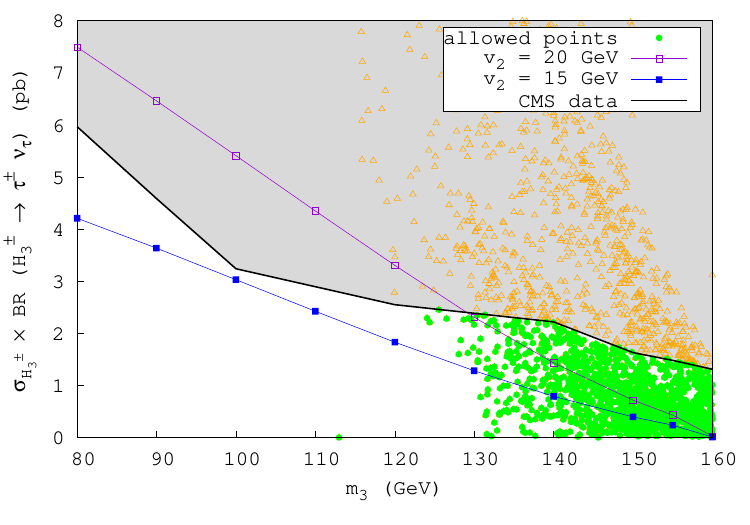} \ \
 \includegraphics[width= 8.2cm]{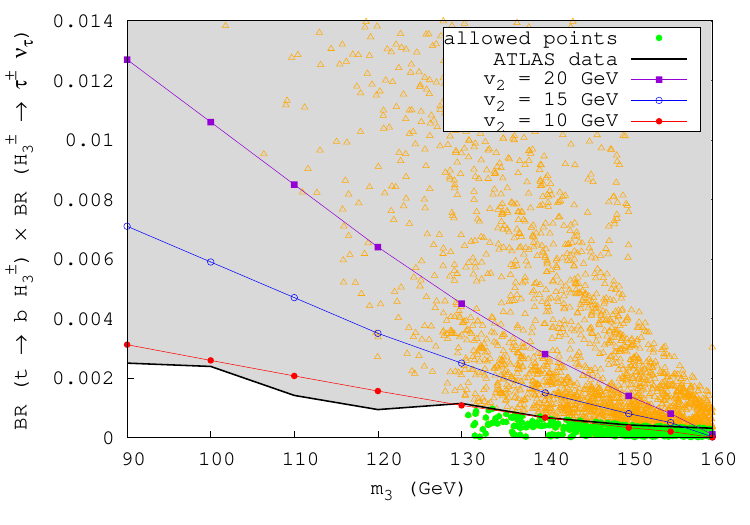}
 \end{center}
 \caption{\small Left : Production cross section times  branching ratio of $H_3^{\pm}$ decaying into $\tau^{\pm}\nu_{\tau}$ as a function of mass $m_3$ in GeV, for $v_2=15,20$ GeV. The black line represents the limit from CMS experiment \cite{CMS:2019bfg}.\\ 
 Right : Branching ratio of the top quark decaying into $H_3^{\pm}$ and $b$ times the branching 
 ratio of $H_3^{\pm}$ decaying into $\tau^{\pm}\nu_{\tau}$ as a 
 function of mass $m_3$ in GeV, for $v_2=10,15,20$ GeV. The 
 black line represents the limit from ATLAS experiment 
 \cite{ATLAS:2018gfm}.\\ 
 The red, blue and violet lines correspond 
 to $v_2=10,15,20$ GeV respectively. The shaded region is 
 excluded by the experimental data. 
 All the coloured points (green and orange) are allowed by the theoretical constraints. 
 The orange and green points are forbidden and allowed by the corresponding experimental data respectively. }
 \label{fig:taunu}
 \end{figure}

\begin{figure}
 \begin{center}
 \includegraphics[width= 7.5cm]{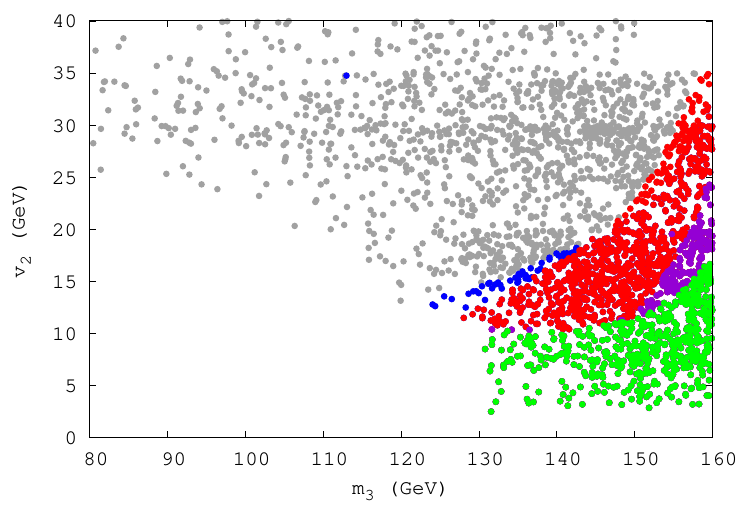} \ \ 
  \includegraphics[width= 7.5cm]{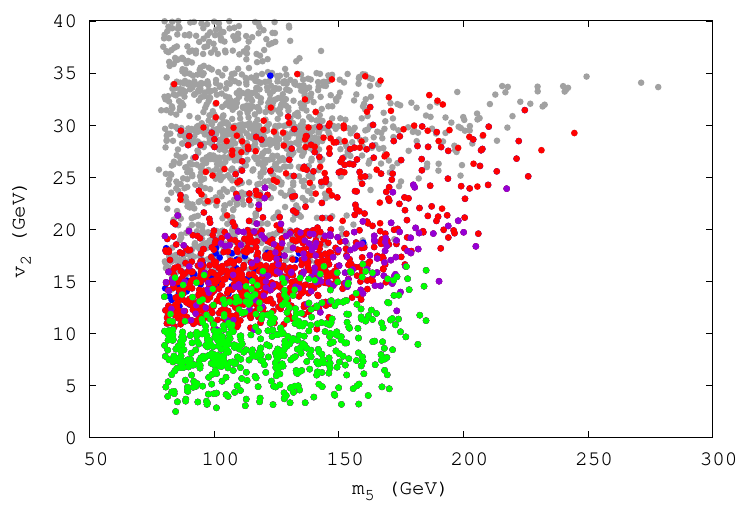} \\
 \includegraphics[width= 7.5cm]{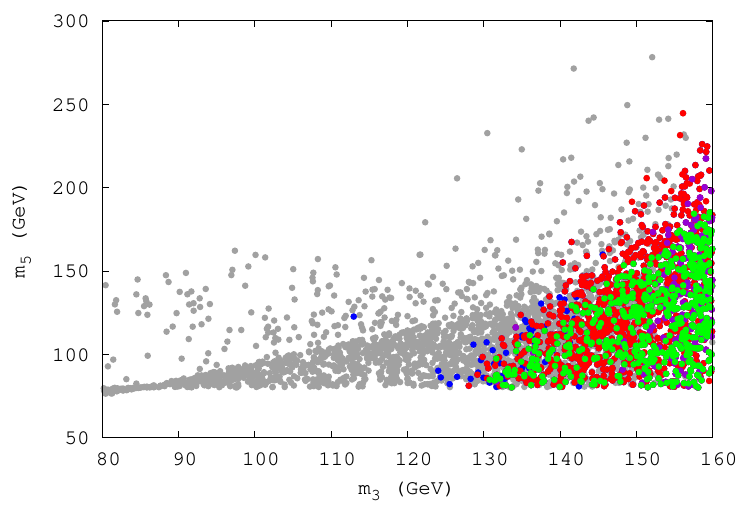} \ \ 
  \includegraphics[width= 7.5cm]{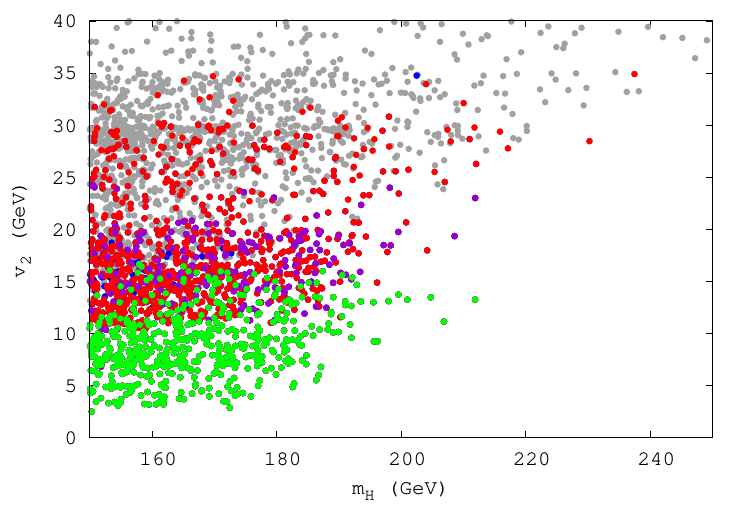} \\
 \includegraphics[width= 7.5cm]{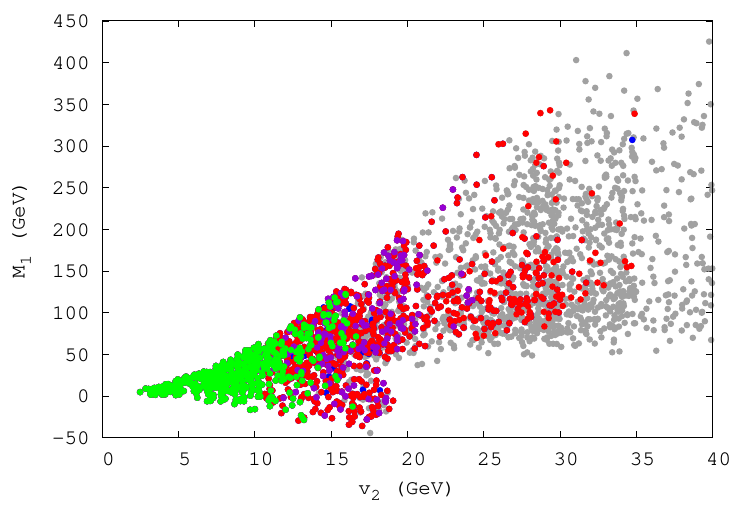} \ \ 
  \includegraphics[width= 7.5cm]{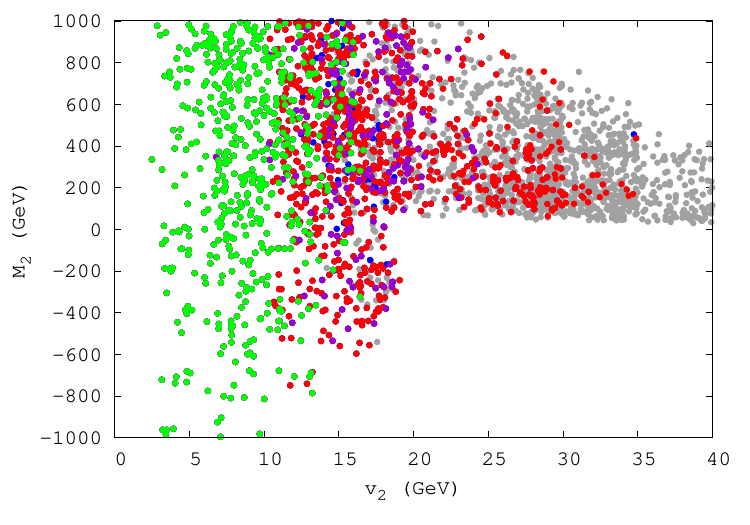} \\
 \includegraphics[width= 7.5cm]{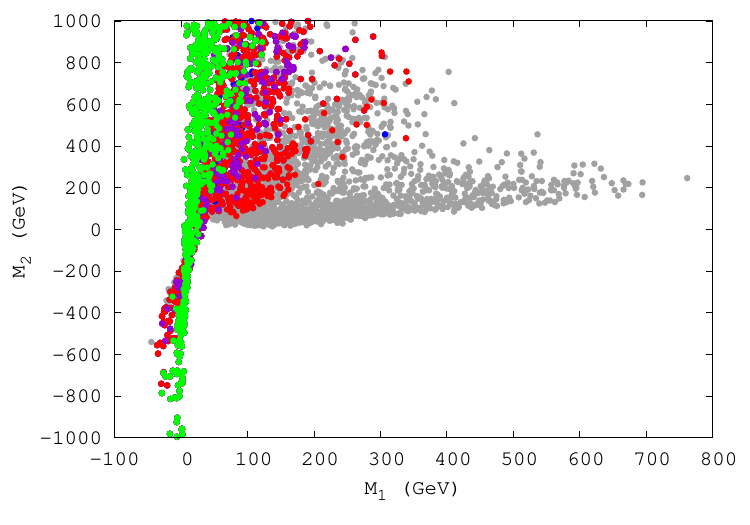} \ \ 
  \includegraphics[width= 7.5cm]{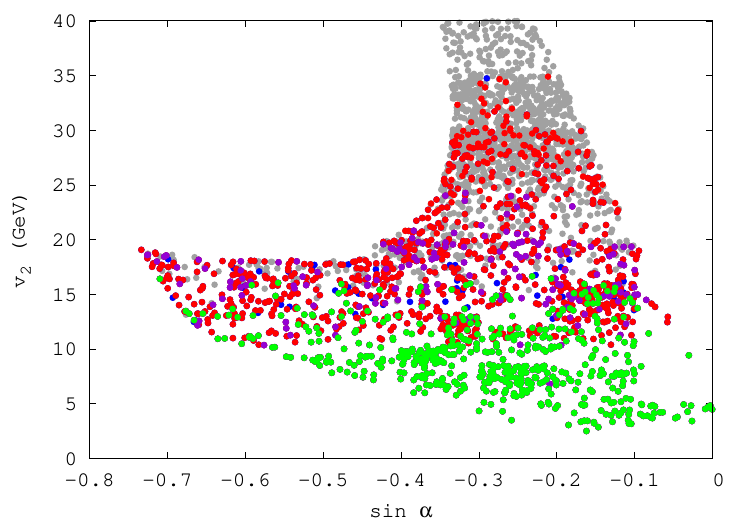} \\
 \end{center}
 \caption{\small Allowed parameter space in the $m_3-v_2$, 
 $m_5-v_2$, $m_3-m_5$, $m_H-v_2$, $v_2-M_1$, $v_2-M_2$, 
 $M_1-M_2$, $\sin{\a}-v_2$ plane. All points are allowed from theoretical constraints 
 as well as LHC data at $\sqrt{s}=13$ TeV. 
 The gray points are excluded from all the $H_3^{\pm} \rightarrow cs, \tau\nu$ data of CMS and ATLAS experiments. 
 The blue points are excluded by $H_3^{\pm} \rightarrow \tau\nu$ ATLAS, $H_3^{\pm} \rightarrow cs$ CMS as well as ATLAS data. 
 The red points are excluded by $H_3^{\pm} \rightarrow cs, \tau\nu$ ATLAS data. 
 The violet points are excluded by $H_3^{\pm} \rightarrow \tau\nu$ ATLAS data. The green points are allowed by all the $H_3^{\pm} \rightarrow cs, \tau\nu$ CMS and ATLAS data. }
 \label{fig:final}
 \end{figure}

\begin{figure}
 \begin{center}
 \includegraphics[width= 9.5cm]{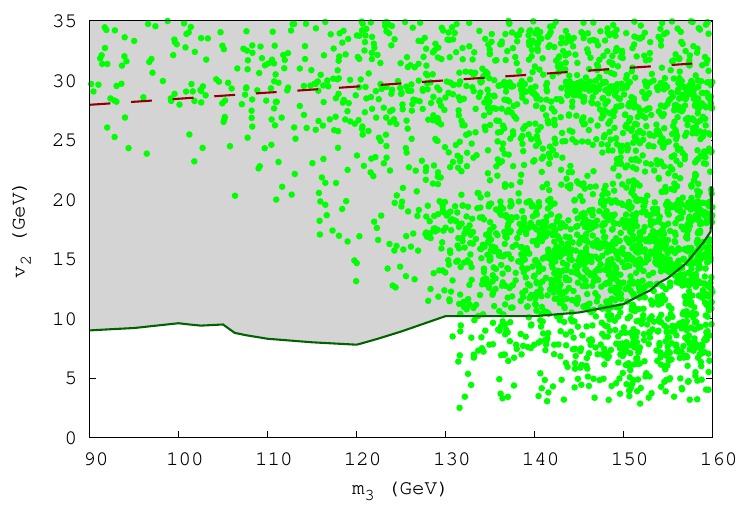} \ \
 \end{center}
 \caption{\small The allowed $m_3 - v_2$ parameter space. 
 The theoretical constraints and the LHC Higgs signal data at $\sqrt{s} = $13 TeV allow all the green points. 
 The region below the dashed line (taken from \cite{Hartling:2014aga}) are allowed by the $b \rightarrow s \gamma$ data. 
 The gray shaded region is forbidden by the $H_3^{\pm} \rightarrow \tau\nu$ ATLAS data. 
 The solid dark-green line corresponds to the $H_3^{\pm} \rightarrow \tau\nu$ ATLAS data for the parameter choice stated in the Eq.\ (\ref{eq:BP}). 
 The green points bellow the solid line are allowed by the theoretical constraints, LHC Higgs signal strengths at LHC $\sqrt{s} = $ 13 TeV, $b \rightarrow s \gamma$ data, $H_3^{\pm} \rightarrow cs, \tau\nu$ ATLAS and CMS data.}
 \label{fig:GM_m3_v2_combined}
 \end{figure}

\begin{figure}
 \begin{center}
 \includegraphics[width= 8.2cm]{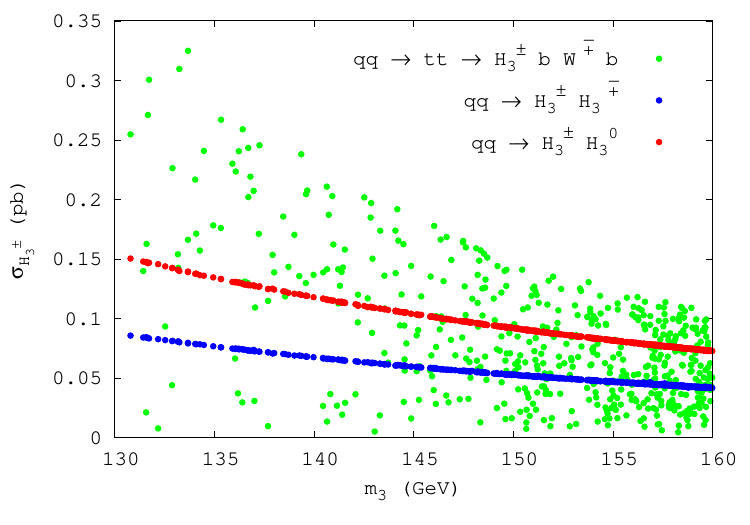} \ \
 \includegraphics[width= 8.2cm]{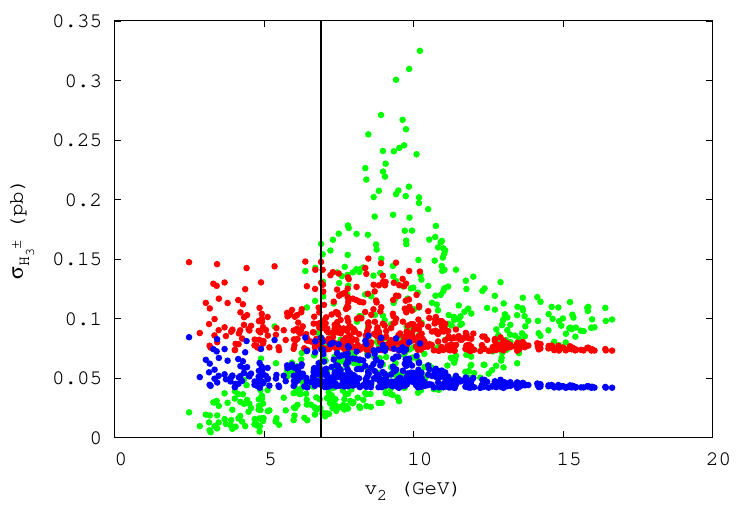}
 \end{center}
 \caption{\small Production cross-sections $\sigma_{H_3^{\pm}}$ for $(i)\, qq \rightarrow tt \rightarrow H_3^{\pm} b W^{\mp} b$ (green points), $(ii)\, qq \rightarrow H_3^{\pm} H_3^{\mp}$ (blue points), $(iii)\, qq \rightarrow H_3^{\pm} H_3^0$ (red points) as a function of (Left) $m_3$ in GeV and (Right) $v_2$ in GeV. Color code is the same for the two plots. 
 The black vertical line in the right figure at $v_2 \approx 7$ GeV separates two regions where the production cross-section of $qq \rightarrow tt \rightarrow H_3^{\pm} b W^{\mp} b$ is lower (left region) or greater (right region) than that of the other two processes.}
 \label{fig:DrellYan}
 \end{figure}

First, we consider the theoretical constraints given in the Eqs. (\ref{eq:unitarity}, \ref{eq:stability}) that restrict 
the quartic couplings $\l_i$s of the scalar potential, which 
are shown by the blue points in the Fig.\ (\ref{fig:overlapped}). 
We fixed the upper limit of $m_3$ at $160$ GeV. 
The expression of $m_3$ given in the Eq.\ (\ref{eq:m3m5}) shows that this puts a limit on $M_1$ in turn. 
We varied the other trilinear co-efficient $M_2$ in a selected range of [-1000,1000] GeV. 
Also, the triplet VEV $v_2$ is varied between [$0,60$] GeV and the doublet VEV $v_1$ is varied accordingly (see Eq.\ \ref{eq:v1v2}). 
So, one can say in some sense that besides the quartic couplings $\lambda_i$s, the trilinear co-efficients $M_{1,2}$ and the VEVs $v_{1,2}$  also acquire limits. 
Now, the expressions of $m_5$ (Eq.\ \ref{eq:m3m5}) and $m_H$ (Eq.\ \ref{eq:mhmH}) show that these masses depend on the $\lambda_i$s, $M_{1,2}$ and the VEVs. Therefore, fixing these parameters also puts some limits on $m_5$ and $m_H$ and this can be seen from the Fig.\ (\ref{fig:overlapped}). 

Next, we consider the LHC data at $\sqrt{s}=$13 TeV, which are 
depicted by the violet points in the plots of the 
Fig.\ (\ref{fig:overlapped}). 
No limit is offered on the masses and the trilinear coefficients 
$M_1$ and $M_2$ by the LHC data. 
But, we get a prominent allowed region in the $\sin{\a}-v_2$ 
plane from these data, which is mainly due to the inclusion of 
the charged scalars in the loop for the decay channel 
$h\rightarrow \gamma \gamma$. 
If the contribution of the charged scalars $H_3^{\pm},\, H_5^{\pm},\, {\rm and}\, H_5^{\pm\pm}$ are not considered in the $h\rightarrow \gamma \gamma$ decay, then we get the red points in addition to the violet points in the Fig.\ (\ref{fig:overlapped}) which obey the LHC data. 
Mainly negative values and some small positive values ($<0.15$) 
of $\sin{\a}$ are admitted by the LHC data. 
For $v_2>10$ GeV, no positive values of $\sin{\a}$ is allowed 
and for $v_2>20$ GeV, a further narrow region is allowed in 
this plane.

It should be clear from the above discussion, that, the  theoretical constraints or the LHC data alone cannot curb the 
parameter space, and hence we are to consider both of them 
simultaneously, which we show in the plots of the 
Fig.\ \ref{fig:overlapped} by the 
dark-green points. 
The plots in the Fig.\ \ref{fig:overlapped} clearly admit that, 
though the limits on the masses ($m_H, m_3, m_5$) and the 
trilinear couplings ($M_1, M_2$) are mainly arising due to 
the adoption of the theoretical constraints, the limits on 
the triplet VEV $v_2$ and $\sin {\a}$ are direct consequence 
of the application of the LHC data. 

%
\section{Results}
%
We have already mentioned that, in the GM model, there are two 
pairs of singly charged scalars; $H_3^{\pm}$ coming from 
the triplet and $H_5^{\pm}$ coming from the quintuplet. 
But, due to its pure triplet origin, $H_5^{\pm}$ do not couple 
to the SM fermions, while $H_3^{\pm}$ couple to the fermions 
via the doublet mixing, which can also be observed from the 
Eq.\ (\ref{eq:fields}). 
In this paper, we are interested only in fermionic decay of the 
singly charged scalars, and therefore, we will talk about the 
branching ratios of $H_3^{\pm}$. 
Since we focus only on the decays of the light charged scalar, 
in this paper, the production process of $H_3^{\pm}$ we consider 
is through the decay of t (or $\bar{t}$) to $H_3^+b$ 
($H_3^-\bar{b}$). 

First, in the Fig.\ \ref{fig:BRtTOh3pb} we depict the branching 
ratio (BR) of $t\rightarrow H_3^+b$ for varying $v_2$ and 
show that, for a particular mass of $H_3^+$, BR increases with 
increasing $v_2$, while it decreases with increasing mass at 
any fixed triplet VEV. 
Here, we consider three values of $v_2$; $10, 15, 20$ GeV. 
Though the BR of the decay $t\rightarrow W^+b$ is maximum, 
that of $t\rightarrow H_3^+b$ is still sizeable in the GM model. 
In the Fig.\ \ref{fig:h3pBR}, we show the BR of $H_3^{\pm}$ 
as the function of mass $m_3$. 
We only depict the decays of $H_3^{\pm}$ for which the 
corresponding BR is greater than $0.0001$. 
To generate these plots (Fig.\ \ref{fig:BRtTOh3pb}, 
\ref{fig:h3pBR}), we choose, 
\be
m_H = 200\,~\text{GeV}\,,\,\,m_5 = 80\,~\text{GeV}\,,\,\,
\sin{\a} = -0.2\,,\,\,M_1 = M_2 = 100\,~\text{GeV}\,,\,\,
\label{eq:BP}
\ee
such that, the selection of the parameters satisfy the 
theoretical constraints given in the Sec.\ \ref{constraints}. 
To plot the Fig.\ \ref{fig:h3pBR}, we choose a particular value 
of the triplet VEV $v_2=$20 GeV, as the nature of this plot is 
independent of the value of $v_2$. 
The plots in this Fig.\ \ref{fig:h3pBR} show that, the BR is maximum for the 
process $H_3^{\pm}\rightarrow cs$ followed by the process 
$H_3^{\pm}\rightarrow \tau^{\pm}\nu_{\tau}$ with an approximate 
value of $0.57$ and $0.39$ respectively. 
Therefore, we are interested in probing the parameter space 
of the GM model for the light (within the mass range $80-160$ 
GeV) singly charged scalar $H_3^{\pm}$ produced by the mechanism 
$t\rightarrow H_3^{\pm}b$, with the subsequent decays 
$H_3^{\pm}\rightarrow cs$ and $H_3^{\pm}\rightarrow \tau^{\pm} \nu_{\tau}$. 

Light charged Higgs boson decaying via $H^{\pm}\rightarrow cs$ 
has already been searched in CMS \cite{CMS:2020osd} and ATLAS \cite{ATLAS:2024oqu} with $\sqrt{s}=13$ TeV. 
Also, search for light charged Higgs boson in the $H^{\pm}\rightarrow \tau^{\pm}\nu_{\tau}$ decay is alredy carried out 
in ATLAS \cite{ATLAS:2018gfm} and CMS \cite{CMS:2019bfg} with 
$\sqrt{s}=13$ TeV. 
Though we had stringent limits on the VEV $v_2$ as a function of the charged Higgs mass $m_{H^{\pm}}$ from the $b\rightarrow s\gamma$ decay \cite{Hartling:2014aga}, 
we use the experimental results from the charged Higgs decay to provide an upper limit on 
the triplet VEV $v_2$ in the GM model. 

We depict $BR(t\rightarrow H_3^+b)\times BR(H_3^+\rightarrow c\bar{s})$ as a function of the charged scalar mass $(m_3)$ 
for three choices of triplet VEV $v_2$ in the Fig.\ \ref{fig:cs}. 
The left plot is for CMS data and the right plot is for ATLAS data. 
Next, we depict the plots for the decay of $H_3^{\pm}$ into 
$\tau^{\pm}\nu_{\tau}$ but in two different manners to apply 
the bound obtained from the data by ATLAS and CMS, separately. 
In the left plot of the Fig.\ \ref{fig:taunu}, we show 
production cross section $(\sigma_{H_3^{\pm}})$ times branching 
ratio of $H_3^{\pm}$ decaying into $\tau^{\pm}\nu_{\tau}$ as a function of mass $m_3$ in GeV, for $v_2=15,20$ GeV. Following 
\cite{CMS:2019bfg}, we consider $\sigma_{H_3^{\pm}}=2\sigma_{t\bar{t}}BR(t\rightarrow H_3^{\pm}b)(1-BR(t\rightarrow H_3^{\pm}b))$. 
In the right plot of the Fig.\ \ref{fig:taunu}, we show 
BR of the top quark decaying into $H_3^{\pm}b$ times BR of 
$H_3^{\pm}$ decaying into $\tau^{\pm}\nu_{\tau}$ as a function 
of mass $m_3$ in GeV, for $v_2=10,15,20$ GeV. 
Here in the Fig.\ \ref{fig:cs} and Fig.\ \ref{fig:taunu}, the black lines represent the limits from CMS and ATLAS experiments. 
The shaded regions in these plots are excluded by the respective experimental data. 
All the points (orange and green) are allowed by the theoretical constraints and LHC Higgs data, but the orange points are forbidden by the respective experimentally observed data whereas the green points are allowed by the respective experiments. 
Therefore, the set of green points are different for each plot of the Fig.\ \ref{fig:cs} and \ref{fig:taunu}. 
From these decay channels, one can see that ATLAS provides a more 
stringent bound on the triplet VEV than CMS. Also, it can be seen that the $H_3^{\pm}\rightarrow \tau\nu$ decay provides a more stringent bound on $v_2$ than the $H_3^{\pm}\rightarrow cs$ decay from the ATLAS result. 
The $m_3-v_2$, $m_5-v_2$, $m_3-m_5$, $m_H-v_2$, $v_2-M_1$, $v_2-M_2$, $M_1-M_2$, $\sin{\a}-v_2$ plots in the Fig.\ \ref{fig:final} clarify this more clearly. 
All the points in these plots are allowed by the theoretical constraints and the LHC Higgs signal data. 
The gray points are excluded by the $H_3^{\pm}\rightarrow cs, \tau\nu$ data. 
The blue, red, violet, green points are allowed by the CMS $H_3^{\pm}\rightarrow \tau\nu$ data but the blue points are excluded by the other three experiments. 
The red, violet, green points are allowed by CMS $H_3^{\pm}\rightarrow cs$ data but the red points are excluded by the two ATLAS experiments. 
The violet and green points are allowed by the ATLAS $H_3^{\pm}\rightarrow cs$ data but the violet points are excluded by the $H_3^{\pm}\rightarrow \tau\nu$ ATLAS data. 
Only the green points survive all the above mentioned observed data. 
Since all the coloured points except the green points are excluded by the ATLAS $H_3^{\pm}\rightarrow \tau\nu$ data, one can say that among these four experimental data, the $H_3^{\pm}\rightarrow \tau\nu$ ATLAS data impose the most stringent constraints on the GM model parameter space. 

Next, considering the parameter choice given in the Eq.\ (\ref{eq:BP}), we varied $v_2$ for different $m_3$ and calculated $BR(t\rightarrow bH_3^{\pm})\times BR(H_3^{\pm}\rightarrow \tau^{\pm}\nu)$. 
The solid line in the Fig. \ref{fig:GM_m3_v2_combined} corresponds to the observed data from ATLAS for this process in the GM model. 
The dashed line obtained from \cite{Hartling:2014aga} for the $b\rightarrow s\gamma$ experiment was the most stringent bound on $v_2$ for different $m_3$ previously. 
But, the results of this paper show that the most stringent bound on $v_2$ for low $m_3$ ($90-160$ GeV) is now obtained from ATLAS data when $H_3^{\pm}$ decays to $\tau\nu$. 
The shaded region is excluded by the ATLAS data. 
All the green points obey the theoretical constraints and the LHC Higgs signal strength data. 
Therefore, the green points above and below the solid line are forbidden and allowed by the experimental data. 

In addition to the above results, this paper also compares the production cross sections of $H_3^{\pm}$ ($\sigma_{H_3^{\pm}}$) from $t\bar{t}$ and $q\bar{q}$. 
Fig.\ \ref{fig:DrellYan} depicts $\sigma_{H_3^{\pm}}$ for $(i)\, qq\rightarrow tt\rightarrow H_3^{\pm}bW^{\mp}b$ (green points), $(ii)\, qq\rightarrow H_3^+ H_3^-$ (blue points), $(iii)\, qq\rightarrow H_3^{\pm} H_3^0$ (red points) as a function of $m_3$ (left) and $v_2$ (right). 
All the points in these plots are satisfied by all of the above mentioned theoretical constraints and experimentally observed data. 
The vertical solid line in the right plot at $v_2\approx 7$ GeV separates two regions where the production cross-section of $qq \rightarrow tt \rightarrow H_3^{\pm}bW^{\mp}b$ is lower (left region) or greater (right region) than that of the other two production processes. 
The solid line indicating the upper limit on $v_2$ as a function of low $m_3$ in the Fig.\ \ref{fig:GM_m3_v2_combined} shows that the minimum value of the allowed upper limit is always greater than $7$ GeV and hence we can safely consider the production cross section of $H_3^{\pm}$ from the top (anti)quark decay instead of the DrellYan processes as depicted in the Fig.\ \ref{fig:DrellYan}. 

To generate the plots, we first implement the model in 
Feynrules \cite{Alloul:2013bka} to obtain the UFO file required 
to generate events using MadGraph5 (v2.8.2) \cite{Alwall:2014hca}. 
%
\section{Conclusions}

Recent searches at the LHC for light singly-charged scalar 
decaying into either $cs$ or $\tau^{\pm}\nu_{\tau}$ encourage 
us to investigate these decay channels in BSM model with 
triplets. 
We choose the Georgi-Machacek model, where the scalar sector 
of SM is added by a real and a complex triplet, such that, 
the custodial symmetry is preserved at the tree level. 
Besides two doubly charged scalars, there are four singly-
charged scalars in this model, out of which two ($H_5^{\pm}$) 
do not have any couplings with the SM fermions. 
Therefore, we are interested in the production and decay of 
the other two singly-charged scalars ($H_3^{\pm}$). 
Since, we consider only the decay of light charged scalars, 
within the mass range of $80-160$ GeV, the production mechanism 
we contemplate is $t\rightarrow H_3^{\pm}b$, though we compared the production processes of $qq\rightarrow tt\rightarrow H_3^{\pm}bW^{\mp}b$ with that of $qq\rightarrow H_3^+ H_3^-$ and $qq\rightarrow H_3^{\pm}H_3^0$. 
We get an upper bound on the triplet VEV by considering the 
experimental limits provided by ATLAS and CMS at the LHC with 
$\sqrt{s}=13$ TeV, for the decay channels 
$H_3^{\pm}\rightarrow cs$ and 
$H_3^{\pm}\rightarrow\tau^{\pm}\nu_{\tau}$, separately. 
The strongest bound is coming from the $H_3^{\pm}\rightarrow \tau^{\pm}\nu_{\tau}$ decay observed by ATLAS. 
On top of this, we also considered the restrictions on the 
parameter space arising from the theoretical constraints and 
Higgs data. 
We showed that, simultaneous adoption of these constraints 
also curb the parameter space of the GM model. 

\vspace{0.5cm}
{\em{\bf Acknowledgements}} --- The author would like to 
acknowledge the SERB grant CRG/2018/004889 as an honorary fellow without any financial 
support and 
Department of Science and Technology, Government of India for financial support 
through SERB-NPDF scholarship with grant 
no. PDF/2022/001784. 
SG also thanks Prof. Anirban Kundu for useful discussions 
and Dr. Suman Chatterjee for technical help in MadGraph5.


 
\end{document}